\documentstyle[psfig]{mn}

\newcommand{\HI}{H\,{\sc i}}

\newcommand{\kms}{~km\,s$^{-1}$}

\newcommand{\vHI}{$v_{\rm HI}$}
\newcommand{\vsys}{$v_{\rm sys}$}

\newcommand{\Lsun}{~L$_{\odot}$}

\newcommand{\FHI}{$F_{\rm HI}$}
\newcommand{\LB}{$L_{\rm B}$}
\newcommand{\MHI}{$M_{\rm HI}$}
\newcommand{\Mtot}{$M_{\rm tot}$}
\newcommand{\Msun}{~M$_{\odot}$}
\newcommand{\Ho}{H$_{\rm o}$}
\newcommand{\AB}{$A_{\rm B}$}
\newcommand{\note}[1]{$^{({#1})}$}

\newcommand{\etal}{et al.}
\newcommand{\eg}{e.g.} 

\title[ATCA HI Observations of the Peculiar Galaxy IC~2554]
      {ATCA\thanks{The Australia Telescope Compact Array is part of the 
             Australia Telescope which is funded by the Commonwealth of 
	     Australia for operation as a National Facility managed by CSIRO.}
       \HI\ Observations of the Peculiar Galaxy IC~2554}

\author[Koribalski \etal]
{B\"arbel~Koribalski$^1$, Scott~Gordon$^{2,1}$, Keith~Jones$^2$ \\
$^1$Australia Telescope National Facility, CSIRO,
    P.O. Box 76, Epping, NSW 1710, Australia;
    E-mail: Baerbel.Koribalski@csiro.au \\
$^2$Department of Physics, University of Queensland, St. Lucia,
    Brisbane, QLD 4072, Australia \\
}

\date{Received date; accepted date}

\begin{document}

\maketitle

\begin{abstract}
ATCA \HI\ and radio continuum observations of the peculiar southern galaxy 
IC~2554 and its surroundings reveal typical signatures of an interacting 
galaxy group. We detected a large \HI\ cloud between IC~2554 and the elliptical
galaxy NGC~3136B. The gas dynamics in IC~2554 itself, which is sometimes 
described as a colliding pair, are surprisingly regular, whereas NGC~3136B was 
not detected. The \HI\ cloud, which emerges from IC~2554 as a large arc-shaped 
plume, has a size of $\sim30$ kpc, larger than that of IC~2554. The total \HI\ 
mass of the IC~2554 system is $\sim 2 \times 10^9$\Msun, a third of which 
resides in the \HI\ cloud. It is possible that tidal interaction between 
IC~2554 and NGC~3136B caused this spectacular \HI\ cloud, but the possibility 
of IC~2554 being a merger remnant is also discussed. We also detected \HI\ gas 
in the nearby galaxies ESO~092-G009 and RKK~1959 as well as an associated \HI\ 
cloud, ATCA~J1006--6710. Together they have an \HI\ mass of $\sim 4.6 \times 
10^8$\Msun. Another new \HI\ source, ATCA~J1007--6659, with an \HI\ mass of 
only $\sim 2.2 \times 10^7$\Msun\ was detected roughly between IC~2554 and 
ESO~092-G009 and corresponds to a face-on low surface brightness dwarf galaxy. 
Star formation is evident only in the galaxy IC~2554 with a rate of 
$\sim$4\Msun\,yr$^{-1}$.
\end{abstract}

\begin{keywords}
   galaxies: individual (IC~2554, NGC~3136B, ESO~092-G009, RKK~1959) --- 
   galaxies: interactions
\end{keywords}

\section{Introduction}


IC~2554 is a southern galaxy with a peculiar optical appearance. It is also
known as Se~72/1 (S\'ersic 1974), ESO~092-IG012, and IRAS~10075--6647. Fig.~1
shows a deep optical image of IC~2554 (Laustsen, Madsen \& West 1987) revealing
many bright knots as well as a network of dust lanes intersecting the main
body of the galaxy. The optical extent of IC~2554 is 3\farcm25 (see Table~1).
Its irregular and asymmetric appearance has prompted the division into multiple
components, roughly aligned North-South. 
Lauberts \etal\ (1978) describe IC~2554 as a `fox-shaped' system with a complex
spiral pattern. They distinguish three components: two central condensations 
separated by 18\arcsec\ with optical velocities of 1230 and 1310\kms\ and a 
detached arm or satellite to the North (1190\kms). Corwin et al. (1985)
describe IC~2554 as a colliding pair (IC~2554A and B, 0\farcm8 apart) with
plumes. Laustsen et al. (1987) describe IC~2554 as a very disturbed pair of
galaxies with a large number of emission nebulae, indicating high activity of
star formation. Lauberts (1982) and Laustsen et al. (1987) note IC~2554 as 
possibly disturbed by the elliptical galaxy NGC~3136B (ESO~092-G013). But 
IC~2554 has also been classified as a peculiar barred spiral galaxy (de 
Vaucouleurs et al. 1991), where the northern spiral arm appears elongated. 
High-resolution \HI\ observations (presented here) are needed to study the
gas dynamics of this complex system and distinguish between the numerous 
scenarios.

Close to IC~2554 (see Fig.~2) are three known galaxies: the elliptical galaxy
NGC~3136B (1780 \kms), the faint spiral ESO~092-G009 and the galaxy RKK~1959.
For a summary of various optical properties of IC~2554 and the neighbouring
galaxies see Table~1.

\begin{table*} 
\caption{Optical properties of the galaxy IC~2554 and catalogued galaxies 
  within a radius of 90\arcmin. Except for RKK~1959 and ESO~092-G009, only 
  galaxies with known velocities in the range 1000--2200\kms\ are included.}
\begin{tabular}{lccccccccc}
\hline
Name       & Projected &\multicolumn{2}{c}{$\alpha,\delta$(J2000)}
           & Opt. velocity & Morph. & Optical diameter & $b_{\rm t}$ 
	   &  $PA$   & $i$    \\
           & Distance   & [$^{\rm h, m, s}$] & [\degr, \arcmin, \arcsec] 
	   &  [\kms]       & Type   &                  & [mag]       
	   & [\degr] & [\degr]\\
\hline
IC~2554    & ---       & 10:08:50.5 & --67:01:47      & $1251\pm18$\note{1}
           & SB3       & $195\arcsec\times74\arcsec$  & 13.1 & 175 & 64 \\
NGC~3136B  & ~8\farcm2 & 10:10:13.1 & --67:00:18      & $1780\pm31$\note{1}
           & E4        & $98\arcsec\times81\arcsec$   & 13.8 &  30 & 69 \\
RKK~1959   & 13\farcm9 & 10:06:37.3 & --67:06:49      & ---
           & ?         & $20\arcsec\times20\arcsec$   & 17.6 & --- & face-on\\
ESO~092-G009&16\farcm5 & 10:06:02.1 & --67:03:48      & ---
           & I         & $81\arcsec\times13\arcsec$   & 16.3 & 107 & 90 \\
	   & \\
ESO~092-G014&25\farcm9 & 10:10:52.2 & --66:38:51      & $1876\pm35$\note{2}
           & F         & $87\arcsec\times51\arcsec$   & 14.2 & 116 & 85 \\
NGC~3136   & 27\farcm4 & 10:05:48.0 & --67:22:41      & $1713\pm30$\note{1}
           & E3        & $215\arcsec\times161\arcsec$ & 12.3 &  30 & 62 \\
RKK~1920   & 30\farcm1 & 10:04:13.5 & --66:48:43      & $2081\pm84$\note{2}
           & L         & $56\arcsec\times56\arcsec$   & 15.0 & --- & face-on\\
ESO~092-G003&57\farcm0 & 09:59:28.3 & --67:18:18      & $1478\pm207$\note{2}
           & S1        & $101\arcsec\times34\arcsec$  & 14.9 &  96 & 83 \\
RKK~2182   & 66\farcm6 & 10:15:43.3 & --67:55:27      & $1923\pm100$\note{3}
           & S1        & $40\arcsec\times15\arcsec$   & 16.5 & --- & 62 \\
ESO~092-G021&79\farcm4 & 10:21:05.6 & --66:29:31      & $2002\pm52$\note{2}
           & L         & $74\arcsec\times60\arcsec$   & 14.5 &  48 & 58 \\
\hline
\end{tabular}
\flushleft
References: The galaxy position in $\alpha,\delta$(J2000), morphological type,
   optical diameter, and the blue magnitude, $b_{\rm t}$, are taken from 
   Kraan-Korteweg (2000). 
   All position angles ($PA$) are from de Vaucouleurs \etal\ (1991), and 
   inclination angles ($i$) are mean values obtained from the Lyon-Meudon 
   Extragalactic Database (LEDA). The velocity references are as follows:
\note{1} de Vaucouleurs \etal\ (1991);
\note{2} Kraan-Korteweg \etal\ (1995);
\note{3} Fairall, Woudt \& Kraan-Korteweg (1998). \\
Tonry et al. (2001) derive distance moduli of 31.95 and 31.64 for NGC~3136
and NGC~3136B, respectively.
\end{table*}

With a Galactic latitude of $b$ = --9\degr\ both the optical extinction 
(0.884 mag, Schlegel \etal\ 1998) and the density of foreground stars are
relatively high. Despite this, numerous optical studies have been carried 
out. The optical velocities recorded for various positions within IC~2554
range from 1190 to 1474\kms\ (Lauberts et al. 1978, Lauberts 1982, de
Vaucouleurs \etal\ 1991, Fouqu\'e et al. 1992, Sanders et al. 1995). The
value of 1850\kms\ also recorded by Fouqu\'e et al. (1992) appears discrepant.
Both Reif (1982) and Bottinelli et al. (1990) measured an \HI\ systemic 
velocity of $\sim$1378\kms\ and an \HI\ flux density of $\sim$30\,Jy\kms\ 
for IC~2554 (see also Huchtmeier \& Richter 1989).
Aalto et al. (1995) obtained $^{12}$CO and $^{13}$CO measurements of IC~2554
and derive a ratio of $R_{1-0}$ = $^{12}$CO(1--0)/$^{13}$CO(1--0)= 13$\pm$2, 
which is at the high end for normal galaxies ($R_{1-0} \simeq 5 - 15$) but
below the ratio of $R_{1-0} \simeq 20 - 50$ found for mergers (see Taniguchi
et al.  1999).

Assuming \Ho\ = 75\kms\,Mpc$^{-1}$, we adopt a distance of $D$ = 16 Mpc for 
IC~2554 and associated galaxies (1\arcmin\ = 4.7 kpc). Throughout the paper, 
the quoted velocities are in the heliocentric velocity frame.


In the following we describe the observations and data reduction (Section~2),
the \HI\ results (Section~3), and the radio continuum results (Section~4).
The discussion of the galaxies is presented in Section~5, followed by a
comparison of various interaction and merger scenarios in Section~6.
The conclusions are given in Sections~7.

\section{Observations \& Data Reduction}
IC~2554 has been observed with the Australia Telescope Compact Array (ATCA)
from January 1997 through to March 1998. \HI\ line and 13-cm radio continuum
data have been obtained simultaneously using three 12-hour observations with
the 750D, 1.5A, and 6B arrays, plus several shorter observations with the 375
array. The galaxy was also observed in the radio continuum at 6 and 3-cm for
a total of 9 hours using the 1.5D array. For details of the observations see
Table~2.

\begin{table*} 
\caption{Observing parameters.}
\begin{tabular}{lc}
\hline
ATCA configurations    & 375, 750D, 1.5A, 6B (20 and 13-cm) \\
                       & 1.5D (6 and 3-cm) \\
Pointing position      & $10^{\rm h}\,09^{\rm m}\,00^{\rm s}$, 
                         --67\degr\,02\arcmin\,00\arcsec \\
Primary calibrators    & PKS~1934--638 or 0823--500 \\
Secondary calibrators  & PKS~0823--500, 1057--797 or 1036--097 \\
{\bf \HI\ data:}       & \\
Centre frequency       & 1413 MHz \\
Total bandwidth        & 8 MHz \\
No. of channels        & 512   \\
Synthesised beam       & $29\farcs5 \times 28\farcs3$ \\
r.m.s. noise           & $\sim$0.75 mJy\,beam$^{-1}$ \\
{\bf Radio continuum data:}  & \\
Centre frequencies     & 2368/2496, 4800, 8640 MHz \\
Total bandwidth        & 128 MHz \\
No. of channels        & 32   \\
\hline
\end{tabular}
\end{table*}

All data were calibrated with the {\em MIRIAD} software package, unless
indicated otherwise, using standard procedures. The 20-cm data was split
into a radio continuum and an \HI\ line dataset using a first-order linear
fit to the line-free channels. 
The \HI\ channel maps (Figs.~3--5) were made using `natural' weighting of 
the data in the velocity range from 1150 to 2040\kms\ (in steps of 10\kms).
The resulting synthesised beam is $29\farcs5 \times 28\farcs3$, and the 
measured r.m.s. is $\sim$0.75 mJy\,beam$^{-1}$. All data are corrected for
primary-beam attenuation. The \HI\ moment maps (Figs.~6--10) were produced 
with the {\em AIPS} task {\em MOMNT} using a spatial smoothing over 3 pixels
and a flux cutoff of 3\,$\sigma$ (2.25 mJy\,beam$^{-1}$).

All radio continuum maps were made using `natural' weighting, resulting in
synthesised beams of $18\farcs3 \times 18\farcs0$ (20-cm) and $9\farcs4 
\times 9\farcs2$ (13-cm). The 6 and 3-cm radio observations were made at
a slightly different pointing of $\alpha,\delta$ (J2000) = 
10$^{\rm h}$\,08$^{\rm m}$\,55$^{\rm s}$, --67\degr\,02\arcmin\,00\arcsec. 
The synthesised beams are $4\farcs2 \times 3\farcs6$ (6-cm) and $3\farcs4 
\times 2\farcs9$ (3-cm). Edge channels attenuated by more than 50\% were
removed during calibration, leaving a bandwidth of 100 MHz at each frequency.


The measured r.m.s. noise is 0.068 mJy\,beam$^{-1}$ (6-cm) and
0.053 mJy\,beam$^{-1}$ (3-cm), compared to the theoretical noise of 
0.054 mJy\,beam$^{-1}$.

\section{\HI\ Results}
Fig.~6 shows the \HI\ distribution of IC~2554 and its surroundings. We detect
five distinct \HI\ sources. Of these, IC~2554 together with an enormous \HI\
extension to the East is the most remarkable source. Towards the SW, and close 
to the edge of the ATCA primary beam, a compact group of three \HI\ sources 
is seen: the edge-on spiral galaxy ESO~092-G009, the galaxy RKK~1959 
(Kraan-Korteweg 2000) and a new \HI\ source, which is referred to here (by
position) as ATCA~J1006--6710. The fifth \HI\ source, which lies roughly 
between IC~2554 and ESO~092-G009, is also new and in the following referred
to as ATCA~J1007--6659.

Because of their large angular separations and velocity differences, separate
\HI\ channel maps are presented for IC~2554 and its \HI\ extension (Fig.~3),
the compact group near ESO~092-G009 (Fig.~4) and ATCA~J1007--6659 (Fig.~5).

All identified \HI\ sources will be described separately in the following
sections. A summary of the basic \HI\ properties is given in Table~3. To 
calculate the \HI\ mass we adopt a distance of $D$ = 16 Mpc for all \HI\ 
sources in the neighbourhood of IC~2554, although we note that the systemic
velocities, \vsys, differ by several hundred \kms\ (see Table~1 \& 3). 
It is not clear if these velocity differences are due to internal group 
dynamics or reflect the actual distances between the individual galaxies.
In the latter case, IC~2554 and ESO~092-G009 would lie in the foreground, 
followed by RKK~1959 and ATCA~J1006--6710, then NGC~3136B, and most distant
ATCA~J1007--6659.
Tonry et al. (2001) derive a distance modulus of 31.64 for NGC~3136B, which
corresponds to a distance 21.3 Mpc. Unfortunately, no independent distance
estimate exists for IC~2554.

\begin{table*} 
\caption{\HI\ properties of the galaxy IC~2554 and surrounding \HI\ sources.}
\begin{tabular}{lccccccc}
\hline
Name             & \FHI  & \MHI$^{\star}$  & \multicolumn{2}{c}{$\alpha,\delta$(J2000)} 
                 & velocity range & \vHI \\
                 & [Jy\kms] & [10$^7$\Msun] & [$^{\rm h, m, s}$] 
		 & [\degr, \arcmin, \arcsec] & [\kms] & [\kms] \\
\hline
IC~2554          & 29.6  &178.5 &   ---    &     ---    & 1240 -- 1550 & \\
~~-- main body   & 19.3  &116.4 & 10:08:50 & --67:01:52 & 1240 -- 1550 & 1380\\
~~-- tidal cloud & 10.3  & 62.1 & 10:09:30 & --67:00:00 & 1240 -- 1430 & 1310\\
ESO~092-G009     & ~3.8  & 22.9 & 10:06:02 & --67:03:45 & 1380 -- 1520 & 1450\\
RKK~1959         & ~2.8  & 16.9 & 10:06:36 & --67:06:48 & 1570 -- 1650 & 1610\\
ATCA~J1006--6710 & ~1.1  & ~6.6 & 10:06:20 & --67:09:14 & 1710 -- 1730 & 1720\\
ATCA~J1007--6659 & ~0.4  & ~2.2 & 10:07:29 & --66:58:50 & 1970 -- 2000 & 1985\\
\hline
\end{tabular}
\flushleft
$^{\star}$The \HI\ mass was calculated using \MHI\ = $2.356~10^5~D^2$~\FHI\
where \FHI\ is the \HI\ flux density in Jy\kms. \\ We adopt a distance of 
$D$ = 16 Mpc for IC~2554 and the surrounding \HI\ sources.
\end{table*}

\subsection{The Peculiar Galaxy IC~2554} 
The ATCA \HI\ observations reveal a large amount of gas within the disk of
the galaxy IC~2554 as well as an extended \HI\ cloud to the East (Fig.~7).
This broad \HI\ tail or plume extends from the northern end of the disk to 
the North-East and then bends over nearly 180\degr\ counter-clockwise towards
the South-East. The outer debris of the \HI\ cloud appear to end just short 
of the elliptical galaxy NGC~3136B, which was not detected. The \HI\ plume, 
which is substantially larger than IC~2554, has a projected length of at 
least $\sim$6\arcmin\ or 30 kpc in East-West direction. The surrounding 
\HI\ debris possibly suggest an even larger extent at lower column densities.

Judging from the \HI\ distribution alone, the plume resembles that of a 
companion galaxy centred on $\alpha,\delta$(J2000) = 10$^{\rm h}$09\fm5,
--67\degr\,00\arcmin, but no optical counterpart has been identified. The
\HI\ velocity field of the plume has no resemblance to that of a galaxy 
and is more likely due to a planar structure with relatively high velocity
dispersion ($\sim$35\kms, see Fig.~7d), indicating rather turbulent motions
(see also Kregel \& Sancisi 2001).

The IC~2554 system covers a velocity range from about 1240 to 1550\kms\ (see
Fig.~3) and contains a total \HI\ mass of $\sim 2 \times 10^9$\Msun. The \HI\
mass of the galaxy IC~2554 alone, which covers the entire velocity range, is 
$\sim 1.2 \times 10^9$\Msun. The \HI\ plume, which covers the lower velocity
range only (1240--1430\kms), contains a slightly lower \HI\ mass of $\sim 
0.6 \times $10$^9$\Msun, although the separation between plume and galaxy
is not well defined.

A detailed look at the \HI\ velocity field of IC~2554 (see Fig.~8) reveals a
rather regularly rotating, nearly edge-on galaxy. The \HI\ gas in the disk
appears to follow the optical S-shape, which can be described as a bar (length
$\sim$45\arcsec, $PA$ = 30\degr) and two emerging spiral arms. A signature 
of the bar is that the major and minor axes in that region of the disk are
not perpendicular to each other. Towards the northern end of the disk the gas
kinematics change drastically as one component still seems to follow the 
optically extended arm towards the North-West and another component traces 
the beginning of the \HI\ plume in the opposite direction, towards the 
North-East.

The slight central depression in the \HI\ distribution of IC~2554 is not
unusual for spiral galaxies (e.g. M\,51, Rots et al. 1990). However, this 
does not mean there is a lack of neutral gas, as the central area contains 
a substantial amount of molecular gas (see Section~5.3). No \HI\ absorption
is visible in the channel maps.

\subsubsection{Comparison with other \HI\ measurements}
Reif (1982) measured an \HI\ systemic velocity of 1378$\pm$24\kms, a velocity 
width of 400\kms, and a total \HI\ flux density of 29.3$\pm$3.5 Jy\kms\ for
IC~2554 (see also de Vaucouleurs et al. 1991, Huchtmeier \& Richter 1989). 
Bottinelli et al. (1990) obtained very similar values, namely a 20\% velocity
line width of 393$\pm$12\kms\ and a total \HI\ flux density of 30 Jy\kms. Both
measurements were obtained with the Parkes 64-m telescope.

We measure a systemic velocity of $\sim$1380\kms\ and a velocity line width
of $\sim$310\kms\ for IC~2554. Whereas the systemic velocity agrees with that
measured by Reif (1982) and Bottinelli \etal\ (1990), the velocity line width
strongly disagrees. This inconsistency can be resolved as both single-dish 
measurements also include some \HI\ emission from the surrounding galaxies 
at slightly higher velocities. The \HI\ flux densities given by Reif (1982) 
and Bottinelli \etal\ (1990) agree well with our measurement of 29.6 Jy\kms.

IC~2554 is clearly detected in the \HI\ Parkes All-Sky Survey (HIPASS; see 
Staveley-Smith et al. 1996; Barnes et al. 2001). The HIPASS spectrum at the
position of IC~2554 (Fig.~11) reveals \HI\ emission in the velocity range
from $\sim$1220 to 1560\kms, mostly associated with the IC~2554 system, and 
fainter emission from $\sim$1560 to 2020\kms, which can be attributed to the
surrounding \HI\ sources (see Fig.~3).
A point source fit to IC~2554 gives an \HI\ flux density of $\sim$21 Jy\kms,
50\% and 20\% velocity line widths of 240\kms\ and 290\kms, respectively, and a
systemic velocity around 1395\kms. We measure a flux density of $\sim$9 Jy\kms\
for the ESO~092-G009 / RKK~1959 group (1380 to 1800\kms) and $\sim$3 Jy\kms\ for 
ATCA J1007--6659 (1800 to 2020\kms). These flux densities are rather uncertain
and depend strongly on the baseline subtraction and the integrated area.

\subsection{ESO~092-G009 and RKK~1959}
In the following we present the results for  the compact group of three \HI\
sources: the edge-on spiral galaxy ESO~092-G009, the galaxy RKK~1959 and the
\HI\ cloud ATCA~J1006--6710. The \HI\ channel maps are shown in Fig.~4, the
\HI\ moments in Fig.~9, and the \HI\ spectra are displayed in Fig.~12. (Note
that this galaxy group is located close to the edge of the ATCA primary beam,
resulting in some uncertainty of the \HI\ flux density estimates.)

ESO~092-G009 (RKK~1947) is a late-type spiral galaxy with no previously recorded
velocity. Using the ATCA, we detected the galaxy in \HI\ at a systemic velocity
of 1450\kms, which makes it a companion to IC~2554. The \HI\ distribution of
ESO~092-G009 (Fig.~9) resembles an edge-on disk with several faint extensions.
The \HI\ extent of ESO~092-G009 is about $3\arcmin \times 1\arcmin$, slightly
larger than its optical size. We measure an \HI\ mass of only $2.3 \times
10^8$\Msun. The \HI\ velocity field of ESO~092-G009 is fairly regular, covering
a velocity range from 1380 to 1520\kms, and the \HI\ velocity dispersion is 
typical for an edge-on galaxy.

RKK~1959 is a very faint, nearly face-on galaxy close to the Galactic Plane 
($b$ = --9\fdg2, \AB\ = 1.0 mag) with an optical size of $\sim$20\arcsec\
(see Kraan-Korteweg 2000). Its projected distance from ESO~092-G009 is 
4\farcm6 or 21 kpc, if both galaxies are at a distance of 16 Mpc. The \HI\
emission covers a velocity range from $\sim$1570 to 1650\kms\ (see Fig.~5);
the galaxy position angle is about $PA$ = 160\degr.
There is an additional \HI\ component around 1720\kms\ near the position of
RKK~1959, which is likely to be part of the \HI\ source ATCA~J1006--6710.
The \HI\ velocity field and dispersion of RKK~1959 (Fig.~9) are strongly 
affected by this feature, which is discussed below.
We measure an \HI\ mass of $1.7 \times 10^8$\Msun\ for RKK~1959, which is 
slightly lower than that of ESO~092-G009.

The new \HI\ source, ATCA~J1006--6710, lies North of RKK~1959 and ESO~092-G009
at projected distances of 2\farcm9 and 5\farcm7, respectively. The channel
maps best show the extended, but clumpy nature of the \HI\ gas which covers 
a very narrow velocity range of $\sim$1710 to 1730\kms.  The northern most
\HI\ clumps in ATCA~J1006--6710 overlap in position with the receding part
of RKK~1959. Numerous \HI\ clouds within the group (see Fig.~9), particularly
between RKK~1959 and ATCA~J1006--6710, suggest some tidal interaction between
the group members. The clumpy nature of the gas in ATCA~J1006--6710 as well
as the velocity structure suggest it is a tidal \HI\ cloud 
(\MHI\ = $6.6 \times 10^7$\Msun) associated with the galaxy RKK~1959. 
Due to the high density of foreground stars it was not possible to identify
an optical counterpart, although there appears to be a very faint (background)
galaxy at $\alpha,\delta$(J2000) = $10^{\rm h}\,06^{\rm m}\,11^{\rm s}$,
--67\degr\,10\arcmin\,00\arcsec, offset from the centre position
but still within the irregular \HI\ envelope of ATCA~J1006--6710.

The galaxies ESO~092-G009 and RKK~1959 as well as a third source,
ATCA~J1006--6710, comprise a compact group with a total \HI\ flux density of 
7.7 Jy\kms\ and a corresponding \HI\ mass of $\sim 4.6 \times 10^8$\Msun. 
Are the members of this group interacting with each other\,? The \HI\ 
distribution of ESO~092-G009 shows some irregular extensions indicating that
the outer portions of the disk may have been disturbed. 
RKK~1959 also appears disturbed and ATCA~J1006--6710 is most likely an
associated tidal \HI\ cloud. The systemic velocities of ESO~092-G009 and 
RKK~1959 differ by about 160\kms, which could indicate quite a large 
separation. The best indicator for tidal interaction in this group is
possibly the amount of \HI\ cloudlets or debris between the galaxies.

\subsection{The galaxy ATCA~J1007--6659}
The other new \HI\ source, ATCA~J1007--6659, lies roughly between IC~2554 and 
ESO~092-G009 at projected distances of 8\farcm5 and 9\farcm8, respectively.
It is barely resolved in \HI\ and covers a narrow velocity range ($\sim$1970 
to 2000\kms). A very faint optical counterpart is visible in the second 
generation Digitised Sky Survey (DSS\,II) as well as in the blue UK Schmidt 
plates scanned with SuperCOSMOS (see Fig.~10). It appears to be an irregular,
nearly face-on low-surface brightness galaxy with an optical diameter of
$\sim$20\arcsec. ATCA~J1007--6659 contains an \HI\ mass of $2.2 \times
10^7$\Msun. We estimate a systemic velocity of $\sim$1985\kms, which is 
substantially higher than that of IC~2554 (\vHI\ = 1380\kms) and indicates 
that ATCA~J1007--6659 probably lies at a larger distance.

\section{Radio Continuum Results}
Of the \HI\ sources described here, only the peculiar galaxy IC~2554 was
detected in the radio continuum. Figs.~13--15 show the maps at all four
frequencies. We measure the following flux densities: 101 mJy (1413 MHz),
65 mJy (2496 MHz), 29.5 mJy (4800 MHz), and 10.8 mJy (8640 MHz). IC~2554 
is also known as the radio continuum source PMN~J1008--6701 for which 
Wright \etal\ (1994) measured a flux density of 37$\pm$7 mJy at 4.85 GHz. 

The radio continuum emission of IC~2554 is extended at all four frequencies.
It is dominated by a central source with a maximum at 
$10^{\rm h}\,08^{\rm m}$\,50\fs2, --67\degr\,01\arcmin\,54\arcsec, 
coinciding with the center of the barred region, but a few arcseconds
South-West of the previously catalogued galaxy center (see Table~1).
The sensitive 13-cm map (Fig.~14) shows most clearly the extent across the 
central bar of IC~2554 and emission at the beginning of the spiral arms,
possibly following the dust lanes (particularly in the North).

\subsection{Star Formation in IC~2554}
For an individual star-forming galaxy, the star formation rate (SFR) is 
directly proportional to its radio luminosity, $L_{\nu}$, (Condon 1992, 
Haarsma \etal\ 2000). 
The radio continuum emission of IC~2554 arises from the central disk and
two emerging spiral arms. The latter are only visible in the 20 and 13-cm 
radio continuum emission. The total SFR, as estimated from those maps, is
$\sim$4\Msun\,yr$^{-1}$ (see Table~4 for the results at all frequencies).
The 6-cm radio continuum emission was detected only across the central 
bar and the 3-cm emission comes mostly from the nucleus.

The far-infrared luminosity, $L_{\rm FIR}$, of IC~2554, as determined from the 
IRAS 60 and 100\,$\mu$m flux densities, is 8.3~10$^9$\Lsun. This value is low 
compared to $L_{\rm FIR} \ga 10^{11}$\Lsun\ for luminous starburst mergers.
Following Hunter et al. (1986) we derive SFR = 2.1\Msun\,yr$^{-1}$, about half
the value determined from the 13 and 20-cm radio continuum observations.
IC~2554 is clearly resolved at 12 and 25\,$\mu$m, and marginally resolved at
100\,$\mu$m (Sanders et al. 1995).

The nuclear H$_2$ mass, $M_{\rm H_2}$, of IC~2554 is $\sim 0.4 \times 
10^9$\Msun\
which is derived using the standard conversion factor of the CO line intensity
(Aalto et al. 1995) to the molecular hydrogen column density. 

We derive a star formation efficiency (SFE) of $L_{\rm FIR}$/$M_{\rm H_2}
\la 21$\Lsun/\Msun, well above the threshold of 4\Lsun/\Msun\ for star-forming
galaxies. Note that $L_{\rm FIR}$ comes from the whole galaxy disk whereas 
$M_{\rm H_2}$ was only measured in the galaxy centre; our value of 
$L_{\rm FIR}$/$M_{\rm H_2}$ is therefore an overestimate. Loiseau \& Combes
(1993) find nuclear and extranuclear CO(1--0) emission in IC~2554.

\begin{table} 
\caption{Radio continuum flux densities, luminosities and star formation 
         rates for IC~2554.}
\begin{tabular}{cccc}
\hline
 $\nu$  & $F_{\nu}$ & $L_{\nu}$                & SFR     \\
 (GHz)  & [mJy]     & [10$^{20}$ W\,Hz$^{-1}$] & [\Msun\,yr$^{-1}$] \\ 
\hline
 1.413  & 101~~  & 30.94 & 3.7 \\
 2.496  & ~65~~  & 19.91 & 3.6 \\
\hline
 4.800  & ~29.5  & ~9.04 & 2.5 \\
 8.640  & ~10.8  & ~3.31 & 1.4 \\
\hline
\end{tabular}
\end{table}

\section{Discussion}
The peculiar optical appearance of IC~2554 has prompted numerous suggestions
as to its nature (see Section~1). Is IC~2554 interacting with a neighbour
galaxy or is it a merging galaxy pair ? The shape of the extended \HI\
distribution and the gas kinematics as well as the peculiar optical 
appearance of IC~2554 should gives us clues as to its history. 

Toomre \& Toomre (1972) showed that the double tailed morphology of many
galaxy systems is the signature of merging pairs of disk galaxies (see also
Barnes 1988). One tail arises from each progenitor galaxy, on roughly opposing
sides of the remnant. This is particularly true for interacting pairs of
comparable mass (Barnes \& Hernquist 1992, 1996). `The Antennae' (NGC~4038/39)
is a prominent example of this kind of merger (Gordon et al. 2001, Hibbard 
et al. 2001). In contrast to NGC~4038/39, the peculiar galaxy IC~2554 shows
only a single tail or \HI\ plume, mild star-formation, and a regularly rotating
disk. The lack of a second surviving arm suggests some difference in the mass,
structure, or gas content of the progenitor galaxies. For example, it is 
possible that only one of the progenitors was gas-rich, producing a single
gaseous tail. {\em But is IC~2554 a merger remnant\,?} The \HI\ velocity field
appears very regular, in contrast to what we would expect from the violent 
merging of two galaxy disks. The latter is only true for a merger of 
comparable-mass galaxies, which eliminates many of the initial characteristics 
of both progenitors in the formation of a new galaxy (Bendo \& Barnes 2000).
Mergers between galaxies of significantly different masses are less violent,
and for sufficiently large mass ratios, the more massive galaxy may survive
essentially unscathed. The mild star-formation, regular disk kinematics and
one-sided \HI\ plume suggest IC~2554 may be the remnant of a merger between
highly unequal-mass galaxies. 

In the following we first explore the possibility of tidal interactions between 
IC~2554 and (1) the elliptical galaxy NGC~3136B to the East (Section~6.1) and 
(2) the spiral galaxy ESO~092-G009 to the South-West (Section~6.2). In these 
scenarios the northern elongated feature in optical images of IC~2554 (Fig.~1)
is interpreted as a tidally distorted spiral arm. After comparison with similar
systems in the literature (Section~6.3) we also discuss the possibility of
a merger event. In that case the northern elongated feature and the central
region of IC~2554 represent the remnants of the two merging galaxies. The
existence of a rather normally rotating disk in IC~2554, as observed in \HI,
is most remarkable.

\subsection{The elliptical galaxy NGC~3136B}
The elliptical galaxy NGC~3136B is, with a projected distance of 8\farcm2,
possibly the closest neighbour to IC~2554 (see Table~1) and therefore a 
potential interaction partner. Although its distance modulus and systemic 
velocity (1780\kms) suggest a much larger distance than IC~2554.
A tidal interaction is only possible if the two galaxies are relatively 
close, in which case substantial peculiar motions are present in the group.
In the following we assume that NGC~3136B is at the same distance as
IC~2554 ($D$ = 16 Mpc), as noted in Section~3.

No evidence for any disturbance has been found in the optical images of 
NGC~3136B. It was not detected in \HI\ and we derive an upper limit to the 
\HI\ mass of 10$^7$\Msun, assuming a velocity width of 100\kms. McElroy 
(1995) measure an optical velocity dispersion of $\sigma_{\rm v}$ = 165\kms\
for NGC~3136B which gives a total dynamical mass of $2.4 \times 10^{10}$\Msun\
using \Mtot\ = $2.31 \times 10^5 R_{\rm c} \sigma_{\rm v}^2$ with a core 
radius of $R_{\rm c}$ = 3.8 kpc (see Table~1).
The extinction-corrected blue magnitude for NGC~3136B ($b_{\rm tc} \sim 13$ 
mag) corresponds to a blue luminosity of $L_{\rm B} \sim 1.4 \times 10^9$\Lsun. 
This is consistent with the $L_{\rm B} \propto \sigma_{\rm v}^4$ relation
(Faber \& Jackson 1976). The resulting mass-to-light ratio is $M$/\LB\ = 17.

The gas-rich spiral IC~2554 and the gas-poor elliptical NGC~3136B have quite
similar total masses, 2.4 and $3.2 \times 10^{10}$\Msun, respectively,
and would make a spectacular interaction pair if they were close. Preliminary 
results from numerical simulations by Horellou (2002, priv. comm.) suggest
that the \HI\ plume and the large difference in systemic velocities ($\Delta
v \sim 400$\kms) could be the result of a parabolic encounter of the two
galaxies.

We note that to the East of NGC~3136B, and very close to the bright star 
HD~309888, lies another small, uncatalogued galaxy at $\alpha,\delta$(J2000) = 
$10^{\rm h}\,10^{\rm m}\,43^{\rm s}$, --67\degr\,01\arcmin\,00\arcsec.

\subsection{The spiral galaxy ESO~092-G009}
The other neighbour to IC~2554 is the highly-inclined spiral galaxy
ESO~092-G009, at a projected distance of 16\farcm5. The systemic velocities
of both galaxies differ by only 80\kms\ (see Table~3), suggesting a 
separation of $\ga$ 77 kpc.

Although the \HI\ velocity field of ESO~092-G009 is regular, the galaxy 
shows some \HI\ extensions which could be due to tidal interactions. 
The closest neighbour to ESO~092-G009 is the galaxy RKK~1959, which is 
also accompanied by a tidal \HI\ cloud. There is evidence for tidal 
interactions within this group (see Fig.~5), so it is difficult to 
attribute any distortions to the relatively distant galaxy IC~2554.

The total dynamical mass of ESO~092-G009 is $2.6 \times 10^9$\Msun, 
substantially less than that of NGC~3136B ($2.4 \times 10^{10}$\Msun) or
IC~2554 ($3.2 \times 10^{10}$\Msun). RKK~1959 has a total mass similar to
that of ESO~092-G009 if we assume an inclination of 20\degr.


In an interaction involving two galaxies of different sizes/masses (\eg\ 
IC~2554 and ESO~092-G009), the smaller, less-massive galaxy should be the
most disrupted, since it exerts smaller tidal forces on material in its
companion, and it has less ability to resist tidal disruption of its own
material. 
So, here we do have a close interaction partner, in fact a group of galaxies,
but together they only comprise a quarter of the total mass of IC~2554. 
Although this group might have played a role in the interaction scenario, it
is, most likely, not able to produce the \HI\ plume observed in IC~2554.

\subsection{Comparison with similar systems}
The `\HI\ Rogues Gallery' (Hibbard et al. 2001) provides a compilation of 
\HI\ maps of peculiar galaxies and is very useful for the comparison and 
understanding of e.g. interacting and merging galaxies. We find that the 
extended \HI\ distributions of the following systems have some similarity
to that of IC~2554: M\,51 (Rots et al. 1990), NGC~3310 (Kregel \& Sancisi
2001), NGC~2146 (Taramopoulos et al. 2001), NGC~3227 (Mundell et al. 1995)
and NGC~520 (Stanford \& Balcells 1991, Hibbard \& van Gorkom 1996).

\HI\ observations of the galaxy M\,51 (NGC~5194) by Rots et al. (1990) 
reveal a spiral structure in the inner regions and faint \HI\ extensions at 
larger radii (up to 21 kpc) with a very complicated velocity structure.
In addition, they find that from the South a broad \HI\ tail (similar to 
that of IC~2554) curves towards the North and East, over a projected length
of 90 kpc. No strong \HI\ emission is associated with the companion NGC~5195.
Detailed numerical simulations of NGC~5194/5 by Salo \& Laurikainen (2000) 
reproduce many of the observed features; the best result is achieved using a 
multiple-encounter model with a recent passage of the companion. Salo \&
Laurikainen note that the pre-existing spiral arms are washed out by the
tidally triggered spiral arms. 

\HI\ observations of the peculiar spiral galaxy NGC~2146 by Taramopoulos et 
al. (2001) reveal two elongated streams of gas extending up to six Holmberg 
radii. The authors propose that these streams were produced by tidal 
interactions between NGC~2146 and a low-surface brightness companion which 
was destroyed in the encounter and remain undetected at optical wavelengths.
The \HI\ kinematics in the inner part of NGC~2146 are quite distorted although
some resemblance to a rotating disk remains. 

\HI\ observations of the spiral galaxy NGC~3227 by Mundell et al. (1995) 
reveal \HI\ plumes extending $\sim$70 kpc North and $\sim$31 kpc South. These
may be a consequence of interaction with the elliptical companion galaxy
NGC~3226, which is located at the base of the northern plume, but was not 
detected in \HI. Again, the disk of NGC~3227 is, despite the proposed 
encounter, in relatively undisturbed solid-body rotation.

\HI\ observations of the isolated spiral galaxy NGC~3310 (Arp~217) by Kregel
\& Sancisi (2001) reveal a rotating disk and two \HI\ tails resembling the 
tidal tails seen in gravitationally interacting systems or mergers. The large
velocity dispersion of the inner parts of the galaxy suggests a highly 
perturbed disk. According to Kregel \& Sancisi all evidence points to a 
recent merger event, possibly of two galaxies with small but comparable masses,
at least one of which is gas-rich.  They note the remarkable existence of a
disk after such an apparently ``major'' merger event and propose that the 
unsettled disk could be either a newly formed disk with spiral arms and 
on-going star formation or the disturbed disk of one of the progenitors
which has survived the merger and is now undergoing new star formation.

If the massive \HI\ plume emerging from IC~2554 cannot be explained by tidal 
interactions of IC~2554 with the elliptical galaxy NGC~3136B, we need to 
consider the possibility that IC~2554 is a merger remnant. There are other
examples in the literature (see above) which support such a scenario for
IC~2554. An equal mass encounter would have destroyed the progenitor disks, 
but simulations by Barnes (2002) show that in the merger remnant a disk can
be re-build. This would be accompanied by high star formation. Or alternately,
a merger of two galaxies with large mass difference would leave the more
massive galaxy nearly intact whereas the less massive galaxy gets destroyed.
Elmegreen \etal\ ( 1993) consider interactions between galaxies with nearly 
equal masses and with {\em extended gas disks} to explain the formation of
cloud complexes seen in NGC2163/NGC~2207. They also find that large gas disks
lead to an extended gas pool at the end of the tidal tail. This might also 
apply to IC~2554.

\section{Conclusions}
The peculiar galaxy IC~2554 and its surroundings were observed with the ATCA
in both the \HI\ emission line as well as the radio continuum. In addition to
IC~2554 itself, which --- contrary to its peculiar optical appearance --- 
shows a surprisingly regular \HI\ velocity field, we detected a large tidal
\HI\ plume East of IC~2554 as well as numerous neighbouring galaxies and 
\HI\ clouds. Radio continuum emission was only detected in IC~2554, where 
it broadly follows the optical shape of the galaxy.

\HI\ observations of IC~2554 reveal a very broad tail or \HI\ plume with a 
projected length of at least 30 kpc. The extent, shape and mass of the plume,
which shows a relatively high velocity dispersion, as well as the \HI\ debris 
surrounding IC~2554 and the plume indicate that a major merger or interaction
event has happened. This is supported by the peculiar optical appearance of 
the galaxy IC~2554. Optical knots seen throughout the body of the galaxy
suggest on-going star formation, however star formation rates determined from
the radio continuum emission are at $\sim$4\Msun\,yr$^{-1}$ not very high.

The edge-on spiral galaxy ESO~092-G009 is most likely the closest neighbour 
to IC~2554 with a separation of $\ga$77 kpc. Although the inner part of its
velocity field appears regular, the outer \HI\ layer of ESO~092-G009 is clearly
disturbed. Some asymmetry is also seen in the optical appearance. The small 
galaxy RKK~1959, just South-East of ESO~092-G009 has an \HI\ mass of $1.7 
\times 10^8$\Msun\ and a systemic velocity of $\sim$1610\kms. It appears rather
disturbed and is associated with the \HI\ cloud ATCA J1006--6710. All three 
\HI\ sources together build a compact group which appears to be mildly 
interacting with each other.

It is unlikely that the \HI\ plume was created by tidal interactions of IC~2554
with ESO~092-G009 / RKK~1959, because of the large difference in total mass,
which would have disrupted the latter, less-massive system.

We suggest that the \HI\ plume is the result of a major merger or interaction
between the gas-rich galaxy IC~2554 and a less massive companion galaxy. The 
fact that IC~2554 has a regularly rotating disk suggests that either the disk
was not destroyed during the encounter or it was rebuilt through tidal forces.
The gas-poor galaxy NGC~3136B would qualify as the interaction partner only 
if it was at a distance similar to IC~2554. In that case the large difference
in redshifts needs to be explained by peculiar motions within the group. If
NGC~3136B is not the interaction partner, then IC~2554 must be the result of
a major merger where either the unknown intruder was destroyed and the disk
of IC~2554 perturbed or both progenitors merged to form a new disk. The latter
scenario is explored by Barnes (2002) who finds that extended gaseous disks
can settle well after the merger has happened leaving one or more tidal tails 
as the main signature of the merger event. The bar and extended star formation
in IC~2554 are likely consequences of such a merger event. The displacement of
some stellar and gaseous tidal features can be explained naturally by the 
extended gas disks of galaxies; it also puts constraints on the merger 
geometry (Mihos 2001).

\section*{Acknowledgements}
\begin{itemize}
\item We thank S. Laustsen, C. Madsen and R.M. West for their kind permission 
      to reproduce the optical image of IC~2554. 
      Credit: European Southern Observatory (ESO)
\item We acknowledge use of Digitized Sky Survey (DSS) material 
      (UKST/ROE/AAO/STScI) and the SuperCOSMOS Sky Surveys (SSS).
\item This research has made use of the NASA/IPAC Extragalactic Database (NED)
      which is operated by the Jet Propulsion Laboratory, California Institute
      of Technology, under contract with the National Aeronautics and Space
      Administration.
\item HIPASS Data were provided by the ATNF under the auspices of the 
      Multibeam Survey Working Group.
\end{itemize}


\begin{figure*} 
\begin{tabular}{c}
\end{tabular}
\caption{Optical image of the peculiar galaxy IC~2554, reproduced from 
  Plate~88 in Laustsen, Madsen \& West (1987), with kind permission from 
  the authors. The white cross indicates the peak of the radio continuum
  emission.}
\end{figure*}

\begin{figure*} 
\begin{tabular}{c}
\end{tabular}
\caption{Optical image of the galaxy IC~2554 and its surroundings, obtained 
  from the Digitised Sky Survey.}
\end{figure*}

\begin{figure*} 
\begin{tabular}{c}
\end{tabular}
\caption{\HI\ channel maps covering the area around the peculiar galaxy 
  IC~2554, the tidal \HI\ cloud and the elliptical galaxy NGC~3136B. 
  The contours levels are $\pm$2.25, 4.5, 9.0, and 18.0 mJy\,beam$^{-1}$ 
  ($3\sigma\ \times 2^n$ with $\sigma$ = 0.75 mJy\,beam$^{-1}$). 
  The synthesised beam ($29\farcs5 \times 28\farcs3$) is displayed at the
  bottom left, and heliocentric velocities (in \kms) are displayed at the
  top left of each panel.}
\end{figure*}
\setcounter{figure}{2}
\begin{figure*}
\caption{--- continued.}
\end{figure*}
\setcounter{figure}{2}
\begin{figure*}
\caption{--- continued.}
\end{figure*}

\begin{figure*} 
\begin{tabular}{c}
\end{tabular}
\caption{\HI\ channel maps covering the area around ESO~092-G009, RKK~1959 
  and the \HI\ cloud ATCA~J1006--6710. 
  The contours levels are $\pm$3.9, 7.8, 15.6, and 31.2 mJy\,beam$^{-1}$
  ($3\sigma\ \times 2^n$ with $\sigma$ = 1.3 mJy\,beam$^{-1}$). 
  The synthesised beam ($29\farcs5 \times 28\farcs3$) is displayed at the
  bottom left, and heliocentric velocities (in \kms) are displayed at the 
  top left of each panel.}
\end{figure*}
\setcounter{figure}{3}
\begin{figure*}
\caption{--- continued.}
\end{figure*}
\setcounter{figure}{3}
\begin{figure*}
\caption{--- continued.}
\end{figure*}

\begin{figure*} 
\begin{tabular}{c}
\end{tabular}
\caption{\HI\ channel maps for the newly discovered galaxy ATCA~J1007--6659, 
  which lies roughly between IC~2554 and ESO~092-G009. 
  The contour levels are $\pm$3.9, 5.5, and 7.8 mJy\,beam$^{-1}$ 
  ($3\sigma\ \times 2^{n/2}$ with $\sigma$ = 1.3 mJy\,beam$^{-1}$).
  The synthesised beam ($29\farcs5 \times 28\farcs3$) is displayed at the
  bottom left, and heliocentric velocities (in \kms) are displayed at the 
  top left of each panel.}
\end{figure*}

\clearpage

\begin{figure*} 
\begin{tabular}{c}
\end{tabular}
\caption{\HI\ intensity distribution of IC~2554 and the 
  surrounding field. Clearly visible are the galaxy IC~2554 (left),
  ESO~092-G009, RKK~1959 and the two new objects (ATCA~J1006--6710 and 
  ATCA~J1007--6659). The greyscale ranges from 0 to 1.75 Jy\,beam$^{-1}$\kms. 
  The contour levels are --0.02, 0.02, 0.04, 0.08, 0.16, 0.32, 0.64, and 1.28 
  Jy\,beam$^{-1}$\kms. The beam is indicated at the bottom left 
  ($29\farcs5 \times 28\farcs3$). }
\end{figure*}

\begin{figure*} 
\begin{tabular}{cc}
\end{tabular}
\caption{\HI\ moment maps of the peculiar galaxy IC~2554. 
  {\bf(a)} \HI\ intensity distribution overlaid onto an optical image from 
     the Digitised Sky Survey (DSS). The contour levels are $\pm$0.02, 0.04, 
     0.08, 0.016, 0.32, 0.64, and 1.28 Jy\,beam$^{-1}$\kms.
  {\bf(b)} \HI\ intensity distribution; contour levels as in {\bf(a)}.
  {\bf (c)} Mean \HI\ velocity field. The contour levels range from 1275 to 
     1525\kms\ in steps of 25\kms.
  {\bf(d)} \HI\ velocity dispersion. The contour levels are 5, 10, 20, 40, 
     and 80\kms. 
  The beam is indicated at the bottom left of each image 
  ($29\farcs5 \times 28\farcs3$). }
\end{figure*}

\begin{figure*} 
\begin{tabular}{cc}
\end{tabular}
\caption{Close-ups of the \HI\ moment maps.
  {\bf(left)} \HI\ intensity distribution showing only IC~2554 overlaid onto 
     the optical image (see Fig.~1). The contour levels are $\pm$0.02, 0.04, 
     0.08, 0.16, 0.32, 0.64, and 1.28 Jy\,beam$^{-1}$\kms\ (as in Fig.~7). 
  {\bf(right)}: Mean \HI\ velocity field. The contours levels are in steps of 
     25\kms, increasing from 1275 to 1525\kms.
  Regions with \HI\ flux density below 0.04 Jy\,beam$^{-1}$\kms\ 
  have been masked out. ---
  The beam is indicated at the bottom left of each image.}
\end{figure*}

\begin{figure*} 
\begin{tabular}{cc}
\end{tabular}
\caption{\HI\ moment maps showing in closeup ESO~092-G009, RKK~1959, and 
  the newly-discovered object ATCA~J1006--6710. 
  {\bf(a)} \HI\ intensity distribution overlaid onto a 2nd-generation DSS 
     optical image. The contour levels are $\pm$0.04, 0.08, 0.16, 0.32,
     and 0.64 Jy\,beam$^{-1}$\kms.
  {\bf(b)} \HI\ intensity distribution, contours levles as in (a); the
     greyscale ranges from 0 to 1.75 Jy\,beam$^{-1}$\kms.
  {\bf(c)} Mean \HI\ velocity field. The contour levels range
  from 1400 to 1512.5\kms\ in steps of  12.5\kms\ (ESO~092-G009) and
  from 1575 to 1725\kms\ in steps 25\kms\ (others); the greyscale ranges
  from 1350 to 1850\kms.
  {\bf(d)} \HI\ velocity dispersion. The contour levels are 5, 10, 
  20, 30, 40, 50, 60, and 70\kms. }
\end{figure*}

\begin{figure*} 
\begin{tabular}{c}
\end{tabular}
\caption{\HI\ intensity distribution of the newly discovered galaxy 
  ATCA~J1007--6659 overlaid onto an optical image obtained from the 
  blue UK Schmidt plates. The contour levels are $\pm$0.02, 0.04, 0.08, 
  0.12 and 0.16 Jy\,beam$^{-1}$\kms.}
\end{figure*}

\begin{figure*} 
\begin{tabular}{c}
\end{tabular}
\caption{HIPASS spectrum at the position of the peculiar galaxy IC~2554.
         A zero-th order baseline was fitted to the spectrum outside the 
	 dashed lines; the \HI\ emission was fitted inside the dotted lines 
	 (see Section~5.1). }
\end{figure*}

\begin{figure*} 
\begin{tabular}{cc}
\end{tabular}
\begin{tabular}{ccc}
\end{tabular}
\caption{\HI\ spectra for IC~2554, ESO~092-G009, RKK~1959 and the 
  newly-detected objects ATCA~J1006--6710 and ATCA~J1007--6659. }
\end{figure*}

\begin{figure*} 
\begin{tabular}{c}
\end{tabular}
\caption{20-cm radio continuum image of IC~2554 (contours), overlaid onto
  the optical image (see Fig.~1). The contour levels are $\pm$0.78, 1.10, 
  1.56, 2.2, 3.1, 4.4, 6.2, 8.8, 12.5, 17.6, 35.3, and 50 mJy\,beam$^{-1}$
  ($3\sigma \times 2^{n/2}$). The beam is displayed at the bottom left
  ($18\farcs3 \times 18\farcs0$).  }
\end{figure*}

\begin{figure*} 
\begin{tabular}{c}
\end{tabular}
\caption{13-cm radio ontinuum image of IC~2554 (contours), overlaid onto
  the optical image (see Fig.~1). The contour levels are $\pm$0.2, 0.28, 0.4,
  0.56, 0.8, 1.14, 1.6, 2.3, 3.2, 4.5, 6.4, 9.1, 12.8, 18.1 mJy\,beam$^{-1}$ 
  ($3\sigma \times 2^{n/2}$). The beam is displayed at the bottom left
  ($9\farcs4 \times 9\farcs2$).  }
\end{figure*}

\begin{figure*} 
\begin{tabular}{cc}
\end{tabular}
\caption{
  {\bf (left)} 6-cm radio continuum image of IC~2554 (contours), overlaid onto
  the optical image (see Fig.~1). The contour levels are $\pm$0.20, 0.29, 
  0.41, 0.58, 0.82, 1.16, 1.64, 2.31, 3.27, 4.63, and 6.55 mJy\,beam$^{-1}$
  ($3\sigma\ \times 2^{n/2}$). The beam is displayed at the bottom
  right ($4\farcs2 \times 3\farcs6$). ---
  {\bf (right)} 3-cm radio continuum image of IC~2554 (contours), overlaid onto
  the optical image (see Fig.~1). The contour levels are $\pm$0.16, 0.22, 
  0.32, 0.45, 0.63, 0.89, 1.26, 1.79, 2.53, and 3.57  mJy\,beam$^{-1}$ 
  ($3\sigma\ \times 2^{n/2}$). The beam is displayed at the bottom
  right ($3\farcs4 \times 2\farcs9$).}
\end{figure*}

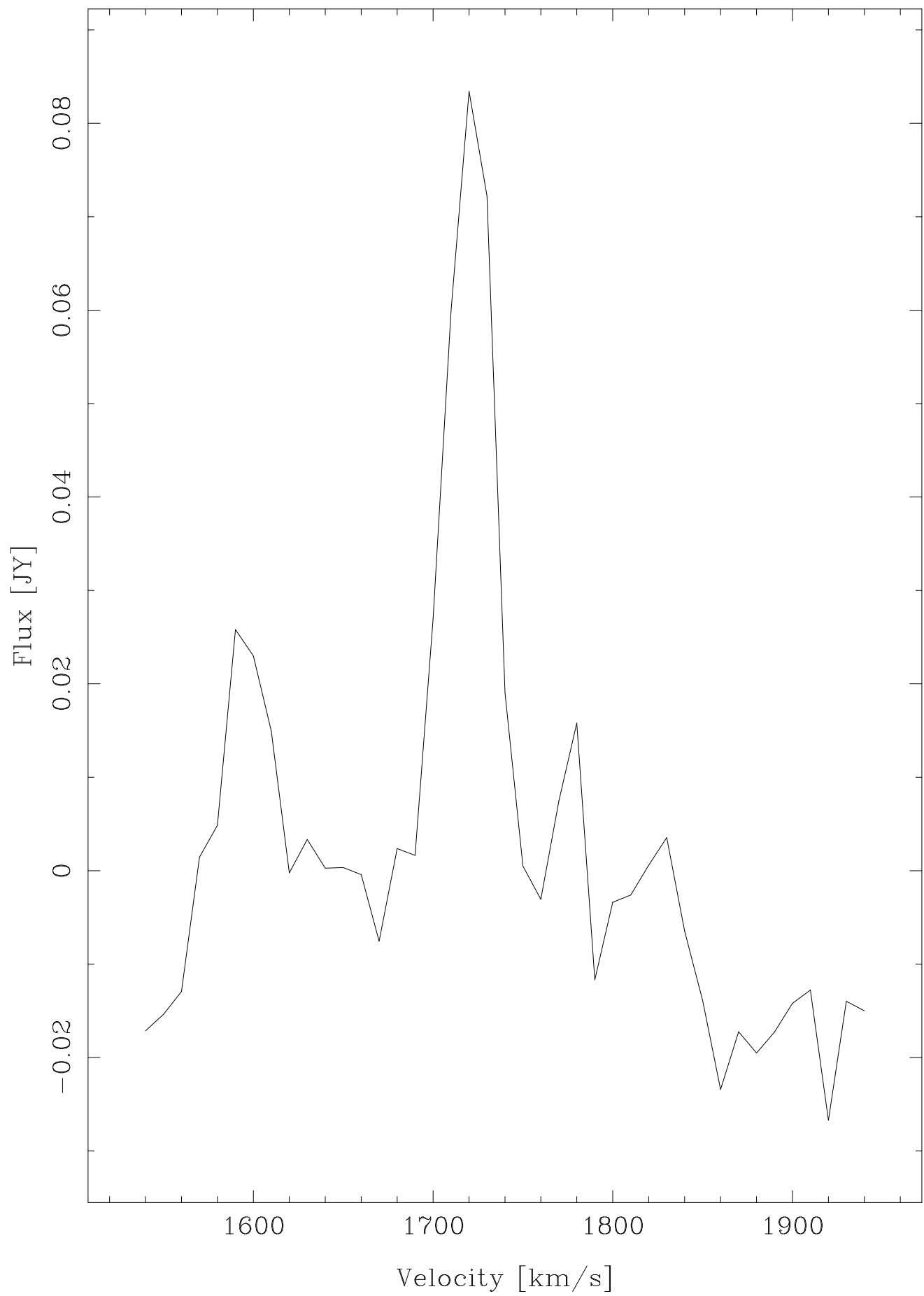
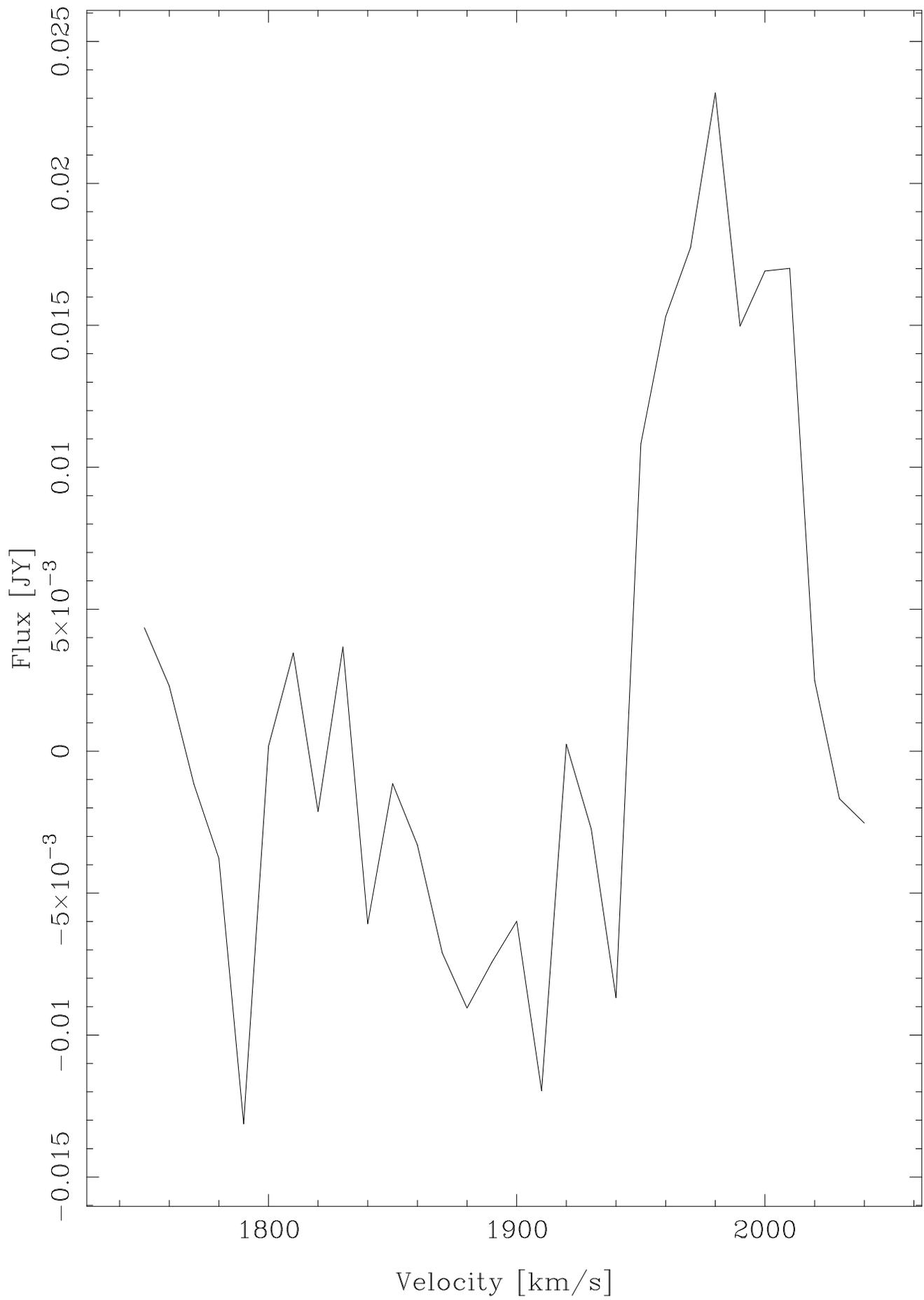
\end{document}